# Photosensitive Strip RETHGEM

V. Peskov[1,2]\*, P. Martinengo[1], E. Nappi[1,3], R. Oliveira[1], G. Paic[2], F. Pietropaolo[4], P. Picchi [5]

[1]CERN, Geneva, Switerland
[2] Inst. de Ciencias Nucleares UNAM, Mexico
[3]INFN Bari, Bari, Italy
[4] INFN Padova, Padova, Italy
[5] INFN Frascati, Frascati, Italy

**Abstract**

An innovative photosensitive gaseous detector, consisting of a GEM-like amplification structure with double-layered electrodes (instead of commonly used metallic ones) coated with a CsI reflective photocathode, is described. In one of our latest designs, the inner electrode consists of a metallic grid and the outer one is made of resistive strips; the latter are manufactured by a screen-printing technology on the top of the metallic strips's grid The inner metallic grid is used for 2-D position measurements whereas the resistive layer provides an efficient spark-protected operation at high gains - close to the breakdown limit. Detectors with active areas of 10x10 and 10x20 cm$^2$ were tested under various conditions including the operation in photosensitive gas mixtures containing ethylferrocene or TMAE vapors.

The new technique could have many applications requiring robust and reliable large-area detectors for UV visualization, as for example, in Cherenkov imaging devices.

Key words: GEM, THGEM, RETHGEM, CsI photocathode, photosensitive gaseous detectors, RICH.

PACS: 29.40

## 1. Introduction

The development of hole-type gaseous photodetectors, (e.g. capillary plates [1, 2], GEM [3], THGEM [4, 5]), combined with solid photocathodes sensitive to UV and even to visible light have been successfully perused by several groups (see for example review papers [3, 6] and references therein). These detectors are proven to be very promising. The main advantage of gaseous detectors operating at atmospheric pressure is the possibility to build large sensitive area (many m$^2$) systems since there are no serious mechanical constrains on the optical-window size. Such large-area photon detectors are attractive for some applications, for example in Cherenkov imaging devices.

In contrast to traditional gas amplification structures, such as parallel-plate or wire type, the hole–type detectors, due to their geometric features, have strong photon-feedback suppression and, in some cases, good ion feedback reduction (see [7] and references therein) ; both are essential for reaching high gas-gains with detectors operating with photocathodes [8]. It is important to cascade a few hole-type structures (several elements operating in tandem), as to reach sufficient gains for efficient single-photon detection. These features allow the photosensitive hole-type structures to reach gas gains of $\geq 10^5$. However, there is a fundamental problem associated with single-photoelectron detection at such high gas gains: the maximum achievable gains $A_{mr}$ of bare hole –type structures (without photocathodes) are governed by the so-called Raether limit [9]:

$$A_{mr}n_0 = Q_{max} = 10^6 \text{-} 10^7 \text{ electrons (1)},$$

where $n_0$ is the number of primary electrons created in the charge-collection region of the detector and $Q_{max}$ is the critical total charge in the avalanche at which the avalanche diverges into a streamer and later, to a discharge.

---

\* Corresponding author: vladimir.peskov@cern.ch

Therefore, at gas gains $\geq 10^5$, any highly-ionizing radioactive background, natural or produced in high-energy physics experiments, if yielding more than 100 primary electrons (e.g. induced by heavily ionizing particles, neutron-interactions, showers and in some cases even minimum ionizing particles), will trigger occasional discharges. The precise value of $Q_{max}$ depends on the detector's geometry and the gas composition, for example $Q_{max}$ is larger in cascaded detectors (due to avalanche-electron diffusion [10]).

When the hole-type structure is combined with photocathodes, the maximum achievable gain $A_{mf}$ can be additionally restricted by the feedback mechanism (see [8] for details) so that

$$A_{mf}\gamma k = 1 \quad (2),$$

where $\gamma$ is the probability of secondary effects (which depends on electric field on the cathode surface and on the gas [8] and k is a coefficient determining what fraction of ions (in the case of the ion feedback) or photons (in the case of the photoeffect) reaches the cathode of the hole-type detector (in the case of wire-type or parallel-type detector usually k~1). Note that very often, particularly in the case of photocathodes sensitive to visible light $A_{mf} < A_{mr}$ [7].

Another drawback raises in case of applications for single-photoelectron detection, where one has to operate the detector at exceptionally high gains, hence at high voltages, entailing that any hole imperfections (tips, dust, dirt) often trigger breakdowns even before reaching the Raether limit [9].

There exist some standard procedures for limiting the destructive power of the discharges in gaseous detectors. For example, in the case of GEMs, it is recommended to segment the electrodes into several independently powered areas and to use protective diodes in the front-end electronics. However, the implementation of these, in the case of large-area devices, e.g. in RICH detectors with sensitive surfaces of several $m^2$, becomes rather complicated and indeed does not guarantee full reliability in long-term exploitation.

To address this problem, we have recently developed what we called a Resistive Electrode Thick GEM (RETHGEM) with electrodes made of resistive materials (instead of traditional metallic ones) [11,12]. The unique property of this detector is that it operates at the same time like a GEM (it can be exploited in a cascaded mode, can operate in badly quenched gases and so on) and like an avalanche resistive-plate chamber (RPC) [13], because it is fully spark-protected. Preliminary tests showed that resistive electrodes can be coated with CsI photosensitive layers and that such detectors may have rather high quantum efficiencies (QE) in the UV range, for example QE ~12.2% was measures at 185 nm in Ne [12].

The technology described in [11, 12] enabled to build only small RETHGEMs, with active area of 3x3 cm$^2$; these early designs had rather large dead-zones (to avoid surface streamers (see [14] for explanations).

To overcome this technical problem, we recently proposed a new design of the RETHGEM with double-layered electrodes: an inner layer consisting of metallic grid or a mesh pattern and the outer one being made of either a uniform or segmented resistive material manufactured by a screen-printing technology [14]. In the latter case, each two neighboring metallic strips were coated with one common resistive strip. This new approach opens the possibility to build large-area RETHGEMs; prototypes with active areas of 10x10 and 10x20cm$^2$ were manufactured and successfully tested with alpha particles and X-rays [14]. This new design also offers better counting-rate charctersics [14].

For the first time, we have studied the feasibility to operate a double-layered RETHGEM as photodetector. The cathodes were either coated with CsI photocathodes or bare-RETHGEMs were investigated in photosensitive gases.

The main focus in these studies has been on the advanced design of the double-layered photosensitive RETHGEM in which each individual metallic strip was coated by its own resistve strip. We call this device strip-RETHGEM or S-RETHGEM. As will be shown below, after optimization of some parameters such as the drift voltage and the gas composition, the S-RETHGEM was able to operate at ten times higher gains than reported earlier in [14]; this allowed not only to detect single photoelectrons with high efficiency, but also to perform 2-D position measurements with UV light.

## 2. Detector Designs and the Experimental Setup

Two designs of photosensitive S-RETHGEMs have been tested in this study: one with strips only on one side of the G-10 plate and the other with strips on both sides. In the latter case, the strips on one surface of the G-10 are perpendicular to the ones on the opposite face.

Fig. 1 schematically shows how the S-RETHGEM is manufactured. First, on the side of the bare G-10 plate (Fig.1a) Cu strips with circular dielectric opening (Fig.1b) were manufactured by a photolithographic technology. The width of the strips, depending on the particular design was 0.7-1 mm, their pitch was 1-1.4 mm and the circle diameter was 0.5-0.8 mm, respectively. Subsequently, the surface of each metallic strip was coated by a screen-printing technology with a 15 μm thick resistive layer, forming the resistive strips (Fig.1c); the gap-width between neighboring resistive strips was 0.1 mm. These double-layered plates were then cured in the oven at temperatures of 200°C followed by holes drilling with a CNC machine - in the middle of each circle opening. Their diameter, depending on the particular design, was in the range of 0.2-0.5 mm. Note that it was very essential to keep the diameter of the drilled holes smaller than the diameter of the uncoated by Cu circles/opening on the metallic strips. This feature provided an efficient spark-protection. S-RETHGEMs with active areas of 5x5 cm$^2$ and 10x10 cm$^2$ have been manufactured following this technique. Finally, the cathode of each S-RETGEM was coated with a 0.35μm thick CsI layer (Fig. 1d).

The inner metallic strips have been used to collect the charge produce by avalanches in holes situated inside each strip. The resistive grid made the detector intrinsically spark protected: in the case of sparks, the resistive layer strongly restricted the current and thus the destructive power.

The experimental setup for the RETHGEM studies is shown in Fig. 2. It consisted of a gas chamber housing a single S-RETHGEM or two S-RETHGEMs operated in cascaded, a Ar(Hg) UV lamp, a monochromator, a lens focusing the light from the lamp to the input slit of the monochromator or directly to the top RETHGEM surface (for position-resolution measurements) and a gas system allowing to flush various gases (He, Ne, Ar or mixture of these with various quenchers including photosensitive vapors: ethylferrocene (EF) or TMAE [8]). For these tests we used rather pure He and Ne (9.9999%), however we did not have the possibility to control the quality of these gases inside the detector's gas chamber.

In some control measurements, the ionization inside the gas has been produced by 5.9 keV photons from a $^{55}$Fe source.

The gain measurements have been performed by two methods: photocurrent measurements (see [15] for details) or measuring the mean value of the pulse amplitudes produced by avalanches with a calibrated charge-sensitive preamplifier [16]. In both cases all metallic strips have been electrically interconnected to the picoampermeter or to the charge-sensitive amplifier, respectively.

In the case of position measurements, several charge sensitive preamplifiers have been connected to the central strips (on which the light of the Hg lamp was focused) whereas the adjacent strips have been grounded via 10M Ω resistors.

The procedure adopted for the quantum efficiency (QE) measurements is the same as described in [12].

**3. Results**

Fig. 3 shows gain vs. voltage curves measured with a single and cascaded S-RETHGEMs in current mode for gains $<10^4$ and in current and pulse mode at gains $>10^4$ in various gases and for two polarities of the electric field in the drift region: a negative one $E_{dr}$= -250V/cm and a positive one (inversed polarity) $E_{dr}$=+250V/cm. The inversion of the electric field in the drift region allows to suppress the contribution of the natural radioactivity and additionally to increase the maximum achievable gains (see the introduction and equation 1). Note that the reversed-field method has been already successfully implemented in cascaded-GEM photon detectors with reflective CsI photocathodes, to suppress charged-particle background in high energy physics experiments [17, 18].

From Fig. 3 one can see that at every polarity of the electric field in the drift region, the gas gains achieved in He- and Ne-based mixtures are an order of magnitude higher than in Ar-based gases. Moreover, in He- and Ne-based gases the S-RETHGEM could operate at 10-20 fold higher gains than in the case of $E_{dr}$= -250V/cm reaching values of $10^5$ and $10^6$ with a single and double S-RETHGEM, respectively. Such high gains allowed detecting single photoelectrons even with one S-RETHGEM.

Note that in Ar- based mixtures the operational voltages were considerably higher that in He- and Ne- based gas mixtures. Thus one can speculate that breakdown in the Ar-based mixture is mainly triggered by the hole imperfections (whereas in He- and Ne-filled detectors the Raether limit could be reached).

Note that due to the low strip capacity and the protective resistive coating, the discharges happening at gains $>10^5$-$10^6$ were rather weak; their energy was almost 10 fold lower than in the case of the ordinary RETHGEMs.

Gain measurements in a pulse-mode (performed at gains $>10^3$ with $^{55}$Fe and at gains $>10^5$ with the UV light) yielded values typically 1.5-2 time higher than in the current mode. We attributed this difference mainly to the charging up effect [14]. Indeed, if the chamber operated in a pulse-mode continuously for 1-2 hours the difference in gain measurements was reduced to 30-50%.

Fig. 4 shows results of the 1-D and 2-D position-resolution measurements performed with single and double S-RETHGEMs. As was mentioned above, in these measurements the UV light from the Ar(Hg) lamp has been focused on the top electrode of the S-RETHGEM (the diameter of the light spot on the cathode surface was~ 0.3 mm) and the avalanche-induced signals have been simultaneously measured from the irradiated strips as well as from several neighboring strips. As one can see from the curves presented in Fig. 4, high-amplitude signals have been measured only from the irradiated strips indicating that a position resolution of about double strip pitch (~2 mm) has been achieved with the single S-RETHGEM and a few times worse (due to the hole misalignment between top and bottom S-RETHGEMs and the diffusion effect) with double S-RETHGEM. One can expect that a better position resolution could be achieved with the S-RETHGEM design having a smaller pitch of holes/strips. For comparison mention that in [19] a position resolution of 0.7 mm FWHM was achieved with double-THGEM irradiated by x-rays and having a separate resistive anode plate (the operation was with electron extraction from THGEM) followed by readout strips of 2 mm pitch.

Among several tested S-RETHGEMs, one has experienced some very mild discharges in Ar+5%$CO_2$ mixture at rather low gains~$10^2$. By identifying the strips at which the discharges occurred and applying a -200V negative voltage on these strips (thus lowering the voltage across the S-RETHGEM in the troubled region), we have been able to operate the remaining surface of the detector at "nominal" gas gains, as shown in Fig. 3. We consider this experience as a possible practical method to operate detectors at high gains even if several holes show defects.

The QE of the CsI coated S-RETHGEM has been measured at 185 nm (see [12] for details) ; it resulted to be ~12% in Ne and ~14% in Ar+10%$CO_2$.

Fig. 5 shows the results of monitoring the QE stability over a period of four months; as one can see from these plots, practically no degradation of the QE has been observed over this period of time. Similar stability was measured with ALICE wire-chamber/CsI photon detectors [20].

In the case of EF and TMAE vapors the S-RETHGEM's sensitivity to the UV light (the S-RETHGEM operated in He-based mixtures) grew almost linearly with the width of the drift gap $D_r$. At $D_r$=5cm the QE was ~1% and ~30% for EF and TMAE; in both cases the detector and the gas system were kept at 27°C. The noticeable difference in the QE is mainly attributed to the very different vapor pressures of EF and TMAE at 27°C [21,22]

## 4. Discussion and conclusions

In this study, for the first time we have tested photosensitive RETHGEMs having double layered strip-electrode structures. This approach allows building large-area detectors. Additionally, the strip geometry offers a very safe operation (small individual strip's capacity and the resistive coatings makes discharges very mild) and the possibility to perform 2-D position measurements. Moreover, in the case of persisting discharges in some particular holes, the strip on which these holes are located could be electrically "disabled" allowing the remaining detector surface to operate at high gas gains.

We have discovered that in He- and Ne- based gas mixtures and in the case of an inverted drift field, S-RETHGEM can be operated at very high gains approaching the Raether limit; this allows detecting single photoelectrons even with a single S-RETHGEM.

The QE of the S-RETHGEMs is sufficiently high for many applications. For example, we are considering using CsI-coated S- RETGEMs in the upgrade of the ALICE RICH detectors, called VHMPIDs [23]. As was shown in [14] the S-RETHEM can operate stably at contng rates <400Hz/cm$^2$, namely at the expected VHPMPID counting rates of the VHMPID detector at ALICE experiment [24]. Another application on which we are working on is an early warning in case of forest fires; to this aim we are testing S-RETHGEMs operating in He + EF gas mixture which allows to detect small flames even in the presence of direct sunlight (see [25] for details).

## Acknowledgements

We thank Amos Breskin for discussions and Miranda Van Stenis for her help throughout this activity. Guy Paic acknowledges the support of the UNAM-DGAPA project IN115808

**Figure captions:**

Fig.1. Steps in the photosensitive S-RETHGEM manufacturing process: a) bare G-10 plate, b) metallic strips with circle opening are manufactured on the G-10 plate by a photolithographic technology, c) resistive strips are manufactured on the top of metallic strips by a screen printing technique and holes are drilled across, d) the top surface of the S-RETHGEM is coated with a CsI layer.

Fig. 2. A schematic drawing of the experimental set up for measuring gain, position resolution and the QE of the S-RETHGEM

Fig.3. Gain vs. voltage curves measured in various gases. Solid symbols-single S-RETHGEM, open symbols –double S-RETHGEM. Triangles- He+EF, squares- Ne, circles- Ar, rhombus - Ar+5%$CO_2$. Red symbols - measurements with single S-RETHGEM and inverted drift field.

Fig.4. Mean signal amplitude vs. strip number measurements at the following conditions: 1, 2-single S-RETGEM, 3, 4 double S-RETGEM. The curve # 1(rhombus), represent the results when middle of the strip #25 of a single S-TETHGEM being irradiated by the UV light, whereas the curve #2 (open squares), shows the results when the UV-light is focused in the region between strips #25 and 26. Curve #3 (triangles) shows the results when the middle of the strip #25 of the top S-RETHGEM was irradiate by the UV light and the signals from the cathode strips of the bottom S-RETHGEM are measured. Curve #4 (stars) the same condition, but the signals from the anode strips are measured

Fig.5. QE vs. time measured in Ar+5%$CO_2$ (rhombus) and Ne (squares) at 1 atm. The Ar(Hg) lamp was kept on and the HV was applied to the detector only during the measurements.

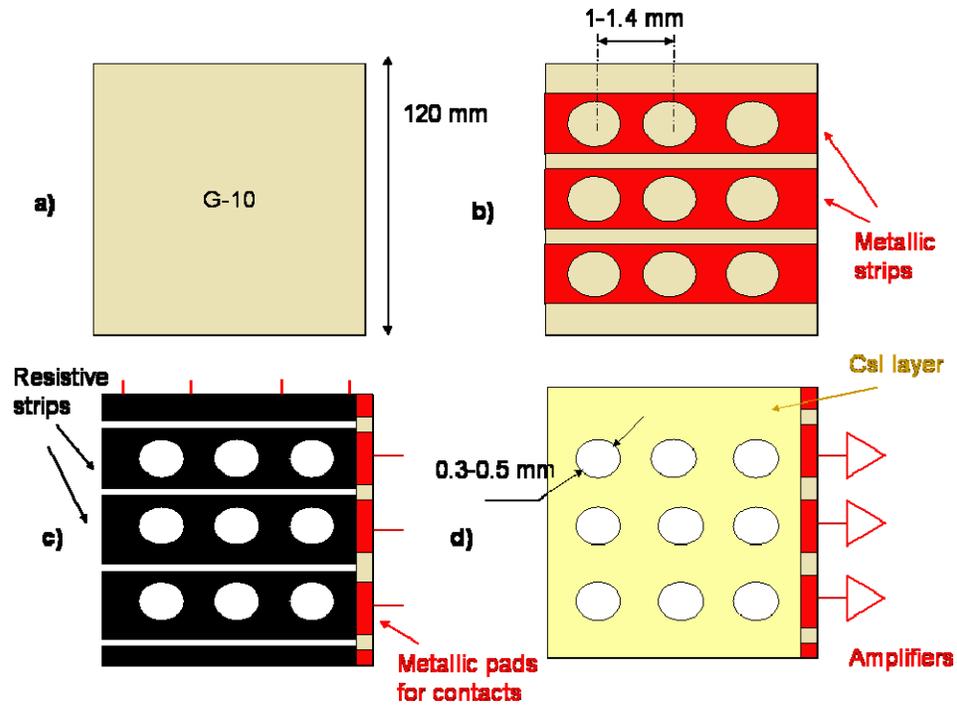

Fig.1

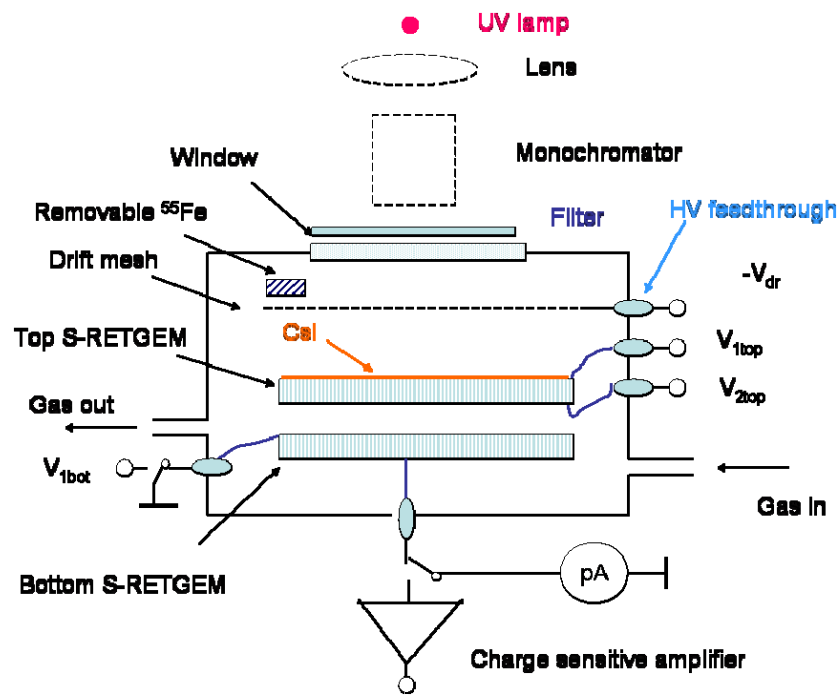

Fig. 2

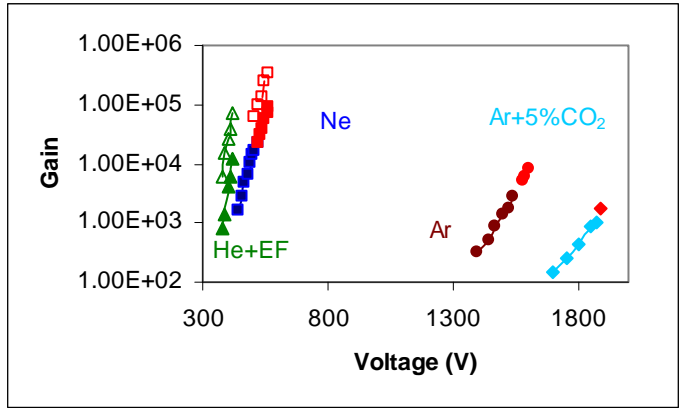

Fig. 3

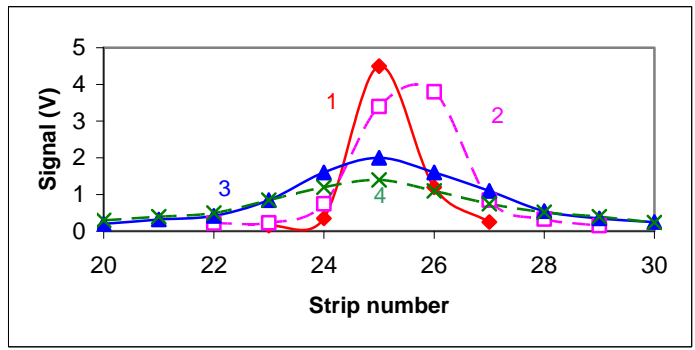

Fig. 4

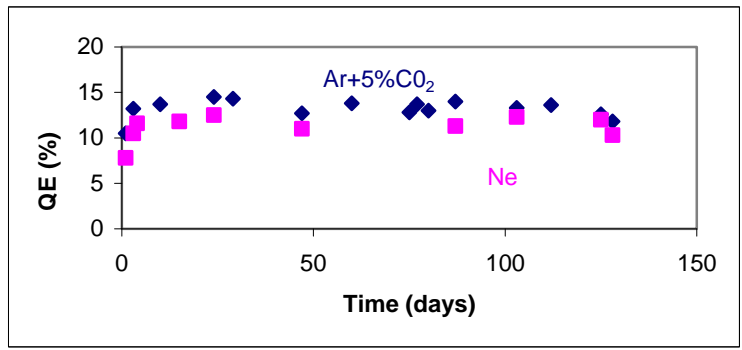

Fig. 5